\documentclass{ws-procs9x6}

\def\al{\alpha}
\def\be{\beta}

\def\et{\eta}

\def\ka{\kappa}
\def\la{\lambda}

\def\rh{\rho}

\def\cL{{\cal L}}
\def\cG{{\cal G}}
\def\cL{{\cal L}}
\def\cE{{\cal E}}

\def\mn{{\mu\nu}}

\def\frac#1#2{{\textstyle{{#1}\over {#2}}}}

\def\prt{\partial}

\def\etal{{\it et al.}}

\def\pt#1{\phantom{#1}}

\def\sb{\overline{s}}
\def\tb{\overline{t}}
\def\ub{\overline{u}}

\newcommand{\beq}{\begin{equation}}
\newcommand{\eeq}{\end{equation}}
\newcommand{\bea}{\begin{eqnarray}}
\newcommand{\eea}{\end{eqnarray}}
\newcommand{\rf}[1]{(\ref{#1})}

\def\etal {{\it et al.}}

\begin{document}

\title{LOCAL LORENTZ-SYMMETRY BREAKING AND GRAVITY}

\author{Q.G.\ BAILEY}

\address{Physics Department, Embry-Riddle Aeronautical University\\
3700 Willow Creek Road, Prescott, AZ 86301, USA\\
E-mail: baileyq@erau.edu}

\begin{abstract}
The lagrangian-based Standard-Model Extension framework 
offers a broad description of possible gravitational effects 
from local Lorentz violation.  In this talk, I review the status of the 
theoretical and phenomenological work in this area.  
The extension of previous results in linearized gravity 
to the nonlinear regime is discussed. 
\end{abstract}

\bodymatter

\section{Introduction}

Although the Standard Model of particle physics and General Relativity (GR)
provide a successful description of all observable physics,
it is widely believed that a unified description exists that 
contains both theories as limiting cases.
This theory remains largely unknown so far and direct 
experimental clues are sparse.

Signals coming from an underlying theory 
that are potentially detectable in sensitive experiments 
include minuscule violations of local Lorentz symmetry.\cite{review} 
The Standard-Model Extension (SME) is a 
comprehensive effective field theory framework that 
describes observable signals of Lorentz violation.\cite{sme}
Much theoretical and experimental work on the SME has 
involved the Minkowski-spacetime limit.\cite{datatables}
Lorentz violation in the gravity sector
has been explored more recently\cite{bk06,b0911,tk} and
experimental analyses have been performed.\cite{expt}
In this talk, 
we discuss the SME framework 
incorporating gravity, with emphasis on spontaneous Lorentz-symmetry breaking.

\section{SME theory}\label{sme theory}

With zero torsion, 
the SME Lorentz-violating couplings linear in the curvature tensor are given by
\beq
\cL = \frac {1}{2\ka} \left( -uR+ s_\mn (R_T)^\mn +t_{\mu\nu\ka\la} C^{\mu\nu\ka\la} \right)+\cL^\prime.
\label{smecouplings}
\eeq
Twenty independent `coefficient fields' are contained in $u$, $s_\mn$, and $t_{\mu\nu\ka\la}$, 
which couple to the Ricci scalar, 
traceless Ricci tensor, 
and the Weyl curvature tensor, 
respectively.
Dynamical terms for these fields are contained in $\cL^\prime$.
Under the assumption of spontaneous Lorentz-symmetry breaking, 
the coefficient fields acquire vacuum expectation values
$\ub$, $\sb_\mn$, and $\tb_{\mu\nu\ka\la}$.

Previous work focused on the limit of linearized gravity 
assuming the metric can be expanded around a Minkowski background,
$g_\mn = \et_\mn + h_\mn$,
where $h_\mn$ are the metric fluctuations.
It is possible to make several assumptions on the dynamics of the 
coefficient fields in order to extract an effective linearized equation for
$h_\mn$ that depends only on the Minkowski metric $\et_\mn$ 
and the vacuum values of the coefficients $\ub$, $\sb_\mn$,  and $\tb_{\mu\nu\ka\la}$
\cite{bk06, s09}.
The linearized ($L$) field equations can then be written in the compact form
\beq
(G_L)_\mn = \ka (T_M)_\mn-\sb^{\al\be} (\cG_L)_{\mu\al\be\nu}.
\label{fe}
\eeq
The coefficient $\ub$ can be removed from the equations at this level as an unobservable scaling, 
while the contribution from the $\tb_{\mu\nu\ka\la}$ coefficients vanishes by a tensor identity.
We use in Eq.\ \rf{fe} the linearized double dual of the Einstein tensor
$(\cG_L)_{\mu\al\be\nu}$.

The field equation \rf{fe} can be shown to satisfy the conservation laws associated 
with local Lorentz symmetry and diffeomorphism symmetry, 
as expected for spontaneous symmetry breaking.
At this level, 
the equations must satisfy the Bianchi identities
$\prt^\mu (G_L)_\mn =0$
and be symmetric in the indices. 
The phenomenology associated with the solutions to Eq.\ \rf{fe},
including the post-newtonian expansion to $PNO(3)$,
is discussed elsewhere\cite{bk06}, 
and experimental limits have been placed on many of the $\sb_\mn$
coefficients\cite{expt}.

It would be of interest to determine the metric component $g_{00}$ to $PNO(4)$, 
which conventionally contains the first terms that exhibit the nonlinearity of GR.
To date, 
the analysis producing Eq.\ \rf{fe} has not been extended to second order in $h_\mn$.
In fact, 
a completely `decoupled' equation to second order, 
involving only the vacuum values of the coefficients, 
and not also their dynamical fluctuations,
may not be obtainable without solving the complete system of equations 
for the metric fluctuations and the coefficient fields.
The process of obtaining general results for the metric $h_\mn$ at higher order 
is therefore likely to be highly model dependent and large in scope.

\section{Models of spontaneous Lorentz-symmetry breaking}\label{models}

Alternatively, one may study specific models of spontaneous Lorentz-symmetry breaking 
and try to generalize the results.
Several types of models exist in the literature
that have a connection to the formalism described above.
One class of models involves an antisymmetric tensor field.\cite{phon}
When nonminimal couplings to gravity are included, 
these models can produce effective $\sb_\mn$ coefficients
and match the form Eq.\ \rf{fe}.

The so-called `bumblebee models' involve a dynamical vector field $B_\mu$ that 
acquires a vacuum expectation value $b_\mu$ 
via a potential term in the lagrangian\cite{ks89}. 
Consider the bumblebee model lagrangian 
\beq
\cL_{B} =
\frac 1 {2\ka} (R + \xi B^\mu B^\nu G_{\mu\nu})
- \frac 14 B^{\mu\nu} B_{\mu\nu} - V+ \cL_{\rm M},
\label{bb}
\eeq
where the field strength is $B_\mn = \prt_\mu B_\nu - \prt_\nu B_\mu$.
The gravitational field equations can be written in the form
\beq
G_\mn = \ka (T_M)_\mn 
+ \ka (B_\mu^{\pt{\mu}\al} B_{\nu\al} - g_\mn B^{\al\be} B_{\al\be} 
- V g_\mn +2 V^\prime B_\mu B_\nu)+\ka \xi (T_\xi)_\mn.
\label{graveqs}
\eeq
The potential energy $V$ is a function of the scalar $X=B^\mu B_\mu - x$,
where $x$ is a constant real number,
thus $V=V(X)$ and $V^\prime = dV/dX$.
The terms proportional to $\xi$ 
are those generated by the nonminimal couplings.

The field equations for the vector field are given by
\beq
D^\mu B_\mn = 2V^\prime B_\nu - \frac {\xi}{\ka} B^\mu G_\mn.
\label{veceqns}
\eeq
The covariant divergence of the left-hand side vanishes identically ($D^\mu D^\nu B_\mn=0$),
which implies a constraint on the right-hand side:
\beq
D^\nu (2V^\prime B_\nu - \frac {\xi}{\ka} B^\mu G_\mn ) = 0.
\label{constraint}
\eeq
Upon expanding around the vacuum values for the metric and vector field,
$B_\mu = b_\mu +\cE_\mu$,
at linear order in $h_\mn$ and $\cE_\mu$, 
the constraint Eq.\ \rf{constraint} becomes $b^\mu \prt_\mu (2 V^\prime) = 0$,
for which the obvious boundary condition choice is 
$(V^\prime)_L = 0$
for both the spacelike and timelike vacuum values.
The linearized limit of this model can be shown to match the form of Eq.\ \rf{fe}.\cite{bk06,s09}

Solving the Eqs.\ \rf{graveqs} and \rf{veceqns} beyond the linearized limit introduces
a number of complexities.
In particular, 
the constraint Eq.\ \rf{constraint} becomes
$b^\mu \prt_\mu (2 V^\prime) = (\xi / \ka) G^\mn D_\mu B_\nu$,
for which the right-hand side has terms that do not vanish at second order in 
the fluctuations.
If the derivative of the potential $V^\prime$ were set to zero (vanishing massive mode condition)
then the right-hand side would also have to be zero,
which is clearly the case for vanishing nonminimal coupling $\xi=0$.
However, 
in order to explore the solutions for nonzero $\xi$,
it appears that we cannot consistently make the choice $V^\prime=0$.

In the post-newtonian limit, 
Eq.\ \rf{constraint} becomes
\beq
b^\mu \prt_\mu (2 V^\prime) = \xi \rh b_j \prt_j U + O(\xi^2),
\label{constraint2}
\eeq
where $\rh$ is the mass density and $U$ is the newtonian potential.
Absent any information about specific boundary conditions, 
we can construct the general solution to Eq.\ \rf{constraint2}
for the spacelike $b_\mu$ case by integrating over the coordinate
$z=\vec x \cdot \hat b$ along the vacuum value direction.

The massive mode combination $V^\prime$ contributes 
as an effective source on the right-hand side of the Einstein equations:
\beq
G_\mn \supset  \ka 2 V^\prime b_\mu b_\nu. 
\label{mm}
\eeq
This source has an intriguing distribution in space for 
an isolated matter source $\rh$.
Outside the matter source where $\rh=0$, 
the right-hand side of \rf{constraint2} vanishes and the massive mode
combination must be independent of one direction (in agreement with other results\cite{bk08}).
On the other hand, there is a finite contribution to the integral of Eq.\ \rf{constraint2}, 
so this remaining source function does not generally vanish as $z \rightarrow \infty$.
The equation for the portion of the metric $g_{00}$ that has $V^\prime$ as its source
is effectively Poisson's equation. 
Detailed solutions including this term, the complete $PNO(4)$ metric, 
and the relevant phenomenology will be presented elsewhere.\cite{ab13}

\end{document}